\documentclass[aps,prl,superscriptaddress,showpacs,floatfix,twocolumn]{revtex4}

\usepackage{graphicx}	

\newcommand{\beq}{\begin{equation}}
\newcommand{\eeq}{\end{equation}}
\newcommand{\beqa}{\begin{eqnarray}}
\newcommand{\eeqa}{\end{eqnarray}}
\newcommand{\bpr}{\begin{problem}}
\newcommand{\epr}{\end{problem}}
\newcommand{\bcent}{\begin{center}}
\newcommand{\ecent}{\end{center}}
\newcommand{\bfig}{\begin{figure}}
\newcommand{\efig}{\end{figure}}
\newcommand{\bpc}{\begin{picture}}
\newcommand{\epc}{\end{picture}}
\newcommand{\barr}{\begin{array}}
\newcommand{\earr}{\end{array}}
\newcommand{\bitm}{\begin{itemize}}
\newcommand{\eitm}{\end{itemize}}
\newcommand{\bright}{\begin{flushright}}
\newcommand{\eright}{\end{flushright}}
\newcommand{\bminip}{\begin{minipage}}
\newcommand{\eminip}{\end{minipage}}
\newcommand{\btab}{\begin{tabular}}
\newcommand{\etab}{\end{tabular}}

\newcommand{\Lmd}{\Lambda}

\newcommand{\lsim}{\mbox{\raisebox{-.3em}{$\;\stackrel{<}{\sim}\;$}}}

\newcommand{\reflef}{(\ref}
\newcommand{\MP}{M_{\rm P}}

\newcommand{\half}{\frac{1}{2}}
\newcommand{\hiroshima}{Graduate School of Science, Hiroshima University, Kagamiyama, Higashi-Hiroshima 739-8526, Japan}
\newcommand{\lmu}{Ludwig-Maximilians-Universit$\ddot{a}$t M$\ddot{u}$nchen, Fakult$\ddot{a}$t f. Physik, Am Coulombwall 1, D-85748 Garching, Germany}
\newcommand{\waseda}{Advanced Research Institute for Science and Engineering,
Waseda University, Okubo, Tokyo, 169-8555, Japan}

\begin{document}

\title{Proposed Laboratory Search for Dark Energy}

\author{Yasunori Fujii} \affiliation{\waseda}
\author{Kensuke Homma} \affiliation{\hiroshima} \affiliation{\lmu}

\date{\today}

\begin{abstract}
The discovery of the accelerating universe indicates strongly the presence
of a scalar field which is not only expected to solve today's version of
the cosmological constant problem, or the fine-tuning and the coincidence
problems, but also provides a way to understand dark energy.
It has also been shown that Jordan's scalar-tensor theory is now going to
be re-discovered in the new lights.
In this letter we propose a way to search for the extremely light scalar
field by means of a laboratory experiment using the low-energy photon-photon
interactions with the quasi-parallel incident beam.
\end{abstract}

\maketitle
%
%
The discovery of the accelerating universe requires us finally to accept
a nonzero cosmological constant, which is, however, smaller 
than the unification-oriented theoretical expectation by as much as 120
orders of magnitude \cite{exp}, known widely as the fine-tuning problem, 
 a part of today's version of the cosmological constant problem.  It
 seems highly remarkable to find that this fine-tuning is  evaded naturally
on the basis of the scalar-tensor theory (STT) invented
first by Jordan \cite{jordan} and now rediscovered with certain new ingredients
included, to implement the scenario of a decaying cosmological constant,
$\Lmd_{\rm obs} \sim t^{-2}$.  The age of the universe $t_0 =1.37 \times 10^{10}{\rm y}$ is re-expressed by
$\sim 10^{60}$ in the reduced Planckian units with $c=\hbar =\MP
(=(8\pi G)^{-1/2})=1$, providing us with an immediate derivation of
 $\Lmd_{\rm obs}\sim 10^{-120}$, or {\em today's $\Lmd$ is
small simply because we are old cosmologically} \cite{cup,ipmurio}.

The scalar field, denoted by $\sigma$, in STT is then expected to fill
up nearly 3/4 of the entire cosmological energy \cite{exp}, known as
dark energy (DE).  Searching for 
this crucial as well as major constituent of the universe by means of
laboratory experiments deserves serious efforts. As we also point out, $\sigma$ is likely massive unlike authentic vector and tensor gauge fields.
According to a simple assumption on the self-energy due to the loops of
ordinary microscopic fields, we suggested an approximate relation
$m_\sigma\sim m_{\rm q}M_{\rm ssb}/M_{\rm P} \sim 10^{-9}{\rm eV}$, in
terms of the $u, d$ quark masses, the supersymmetry-breaking mass-scale
and the Planck mass, respectively \cite{nat,cup}.   This also corresponds to
a {\em macroscopic} distance $m_\sigma^{-1}\sim 100{\rm m}$, though 
we allow for the latitude of a few orders of magnitude. 

Past searches for the scalar force of this kind have been
plagued by its matter coupling basically as weak as gravity \cite{wepv},
inevitably with heavy and huge objects.   This blockade can be removed,
however, by appreciating that the scattering amplitude in which $\sigma$
occurs as a resonance reaches a maximum independent of the interaction
strength, but  with a concomitant narrow width.  Also a resonance as
light as above might be realized only by means of low-energy
photon-photon scattering, unless, as required by the weak
equivalence principle (WEP) \cite{yfms}, $\sigma$ is totally
decoupled from the photons. 
Through detailed analyses of the two-photon systems, 
we propose a novel type of laboratory experiments 
providing a glimpse of DE, anticipating an added building
block $\sigma$ in the extended theory living with the accelerating
universe.  For other theoretical details we  suppress, see our
references \cite{cup,ipmurio,ptpkmyf}.   

\begin{figure}
\includegraphics[width=7.0cm]{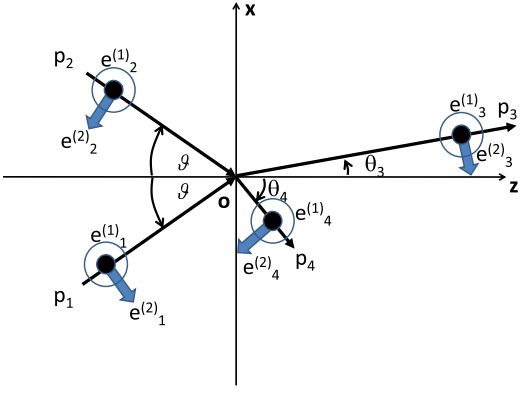}
\caption{
Definitions of kinematical variables.
}
\label{Fig1}
\end{figure}

%
%
For the reasons to be explained shortly, we prefer a special
coordinate frame, as shown in Fig.\ref{Fig1}, in which two 
photons labeled by  1 and 2 sharing the same frequency are 
incident nearly parallel to each other, making an angle $\vartheta$ with
a common central line along the $z$ axis. We define the $zx$ plane
formed by $\vec{p}_1$ and $\vec{p}_2$.  The components of the
4-momenta of the photons are given by $p_1
=(\omega\sin\vartheta,0,\omega\cos\vartheta ; \omega)$ and the same for
$p_2$ but with the sign of $\vartheta$ reversed, and $p_3 =(\omega_3
\sin\theta_3, 0, \omega_3 \cos\theta_3 ; \omega_3)$ and  $p_4$  with
$\omega_3, \theta_3$ replaced by $\omega_4, -\theta_4$, respectively.

The outgoing photons are assumed to be in the same $zx$ plane, 
to be convenient particularly in the $s$-channel reaction, showing
an axial symmetry with respect to the $z$ axis.  The angles $\theta_3$
and $\theta_4$, both positive $<\pi$, are defined also as shown in
Fig.\ref{Fig1}.   This coordinate frame can be transformed from the
conventional CM frame for the head-on collision in the $x$ direction 
by a Lorentz transformation with the relative velocity 
$\beta_z=\cos\vartheta$. 

The conservation laws are
\beqa
0 \mbox{-axis}:&&\omega_3 +\omega_4 = 2\omega, \label{kinm_3}\\
z\mbox{-axis}:&&\omega_3 \cos\theta_3 + \omega_4 \cos\theta_4= 
2\omega \cos\vartheta, \label{kinm_4}\\
x\mbox{-axis}:&&\omega_3\sin\theta_3 =\omega_4\sin\theta_4. 
\label{kinm_5}
\eeqa
For a convenient ordering $0<\omega_4 < \omega_3 <2\omega$, we may choose
$0<\theta_3<\vartheta<\theta_4<\pi$, without loss of generality.  From
\reflef{kinm_3})-\reflef{kinm_5}) we derive $\sin\theta_3 / \sin\theta_4
=\sin^2\vartheta /W$ with
$W=1-2\cos\vartheta\cos\theta_4+\cos^2\vartheta$. 

The differential elastic scattering cross section favoring the
higher photon energy $\omega_3$ is given by \cite{footnote1}
\beq 
\frac{d\sigma}{d\Omega_3}=\left(\frac{1}{8\pi \omega
}\right)^{2}\sin^{-4}\vartheta \left(
\frac{\omega_3}{2\omega} \right)^2 |M|^2,
\label{kinm_13}
\eeq
where $M$ is the invariant amplitude, and $\omega_3 =(\omega
\sin^2\vartheta)/(1-\cos\vartheta \cos\theta_3)$.  
For $\vartheta \ll 1$, we then derive the upshifted frequency $\omega_3
\rightarrow 2\omega$, as $\theta_3\rightarrow 0$, a clear observational
signature, also occurring in the extremely forward direction within the
angle $\vartheta$. 

%
%
The scalar field $\sigma$ may couple to the electromagnetic field with the
effective interaction Lagrangian given by
\beq
-L_{\rm mx\sigma}=(1/4)B\MP^{-1}F_{\mu\nu}F^{\mu\nu}\sigma,
\label{mxelm_1}
\eeq
where, due to the quantum-anomaly-type estimate, the constant
$B$ is proportional to the fine-structure constant \cite{cup, footnote2}.
This interaction term, which has been discussed also from
a phenomenological point of view \cite{bek}, is WEP violating
\cite{yfms}, already in Brans-Dicke's sense \cite{bd}.

We find, for example, 
\beq
<0\:| F_{\mu\nu}|\:p_1,e_1^{(\beta)}\!>=i\left(p_{1\mu}e_{1\nu}^{(\beta)} -
p_{1\nu}e_{1\mu}^{(\beta)}\right),
\label{mxelm_4}
\eeq
giving the two-photon decay rate of $\sigma$ with the mass $m_\sigma$;
\beq
\Gamma_\sigma
=(16\pi)^{-1} \left(
B\MP^{-1}\right)^2 m_\sigma^3,
\label{mxelm_4a}
\eeq
by assuming purely elastic scattering.

The polarization vectors  are given by $\vec{e}_i^{\
(\beta)}$ with $i=1, \cdots,4$ for the photon labels, whereas
$\beta =1,2$ are for the kind of linear polarization, also shown in
Fig.\ref{Fig1}.

In the $s$-channel, the scalar field is exchanged between the pairs
$(p_1, p_2)$ and $(p_3, p_4)$, thus giving the squared momentum
of the scalar field $q_s^2 =\left(p_1+p_2\right)^2 =2\omega^2 \left(
\cos 2\vartheta -1 \right)$ with the metric convention $(+++-)$.

With the type $\beta=1$ for all the photons we find \cite{footnote3} ;
\beq
M_{1111s}=-(B\MP^{-1})^2\frac{\omega^4 \left(
\cos2\vartheta -1\right)^2}{2\omega^2 \left(
\cos2\vartheta -1\right)+m_\sigma^2},
\label{mxelm_7}
\eeq
where the denominator, denoted by ${\cal D}$, is the $\sigma$ propagator.
We note $q_s$ is timelike, unlike $t$- and $u$-channels.
We then make a replacement
\beq
m_\sigma^2 \rightarrow \left( m_\sigma -i\Gamma_\sigma \right)^2 \approx
m^2_\sigma -2im_\sigma \Gamma_\sigma.
\label{mxelm_9}
\eeq
Substituting this into the denominator in \reflef{mxelm_7}), and
expanding around $m_\sigma$, we obtain
\beq
\hspace{-.1em}{\cal D}\approx -2\left( 1-\cos2\vartheta  \right) \left( x+ia
\right),\quad\hspace{-.7em}\mbox{with}\quad\hspace{-.7em} x =\omega^2 -\omega_r^2,
\label{mxelm_10}
\eeq
where
\beq
\omega_r^2 =\frac{m_\sigma^2/2}{1-\cos 2\vartheta },\quad
a=\frac{m_\sigma \Gamma_\sigma}{1-\cos 2\vartheta}.
\label{mxelm_12}
\eeq
Notice that both of $\omega_r^2$ and $a$ are enhanced as $\vartheta
\rightarrow 0$.

Using \reflef{mxelm_4a}) and \reflef{mxelm_12}) repeatedly, we finally
obtain 
\beq
|M_{1111s}|^2 \approx  (2\pi)^2 \frac{a^2}{x^2+a^2},
\label{mxelm_13}
\eeq
from which  we derive $|M_{1111s}|^2_{\omega=\omega_r}=(2\pi)^2$,
a ``large'' value entirely free from being small due to the factor $\MP^{-4}$.
This is an aspect in the efforts to overcome
the weak coupling of gravity, as alluded at the beginning.
We may then ignore non-resonant terms in the $s$-channel and
the whole contribution from the $t$- and $u$-channels.
We still face the weak coupling in the extremely narrow width $a$,
implied by $\MP^{-2}$,  in \reflef{mxelm_4a}) and
the second of \reflef{mxelm_12}).

To cope with this, we apply a process of averaging;
\beq
\overline{|M_{1111s}|^2}=\frac{1}{2\tilde{a}}\int_{-\tilde{a}}^{\tilde{a}}
|M_{1111s}|^2 dx = (4\pi)^2 \eta^{-1}\frac{\pi}{2}\hat{\eta},
\label{mxelm_15}
\eeq
over the range $2\tilde{a}$ of $x \sim \omega^2$, where $\eta
\equiv \tilde{a}/a$, also with $\hat{\eta} = (2/\pi) \tan^{-1}\eta $
reaching the maximum 1 for $\eta\rightarrow \infty$.

Substituting \reflef{mxelm_15}) into  \reflef{kinm_13}) we obtain
\beq
\overline{\left(\frac{d\sigma}{d\Omega_3}\right)}_{1111s}
=\frac{\pi}{8\omega^2}\sin^{-4}\vartheta \left( \frac{\omega_3}{2\omega}\right)^2\eta^{-1}\hat{\eta}.
\label{mxelm_19}
\eeq
For later convenience, we parametrize $\eta$ by
\beq
\eta =\left( \MP/m_\sigma \right)^{2-\gamma}.
\label{mxelm_19a}
\eeq

We also derive $M_{1111s}=M_{2222s}=-M_{1122s}=- M_{2211s}$,
for the only nonzero components.   We are especially interested in
$M_{1122s}$ from the experimental point of view as explained later.

In order to design experiments, we start with the resonance condition,
the first of \reflef{mxelm_12}), by assuming $\vartheta \ll 1$,
\beq
m_{\sigma}/2 \sim \vartheta \omega.
\label{exeq_1}
\eeq
This indicates that experiments have the two adjustable handles for a given
scalar mass scale. Since scanning $\vartheta$
would be much easier than scanning $\omega$, for the following argument,
we assume fixing the incident energy and scanning $m_{\sigma}$ by
changing $\vartheta$. We consider a case of the resonance condition
with $\omega\sim 1$~eV (optical laser) and $\vartheta\sim10^{-9}$
for $m_{\sigma}\sim 10^{-9}$~eV in what follows.

We find it useful to  approximate $a$ in \reflef{mxelm_12}) by
\beq\label{exeq_2}
a \sim \kappa( m_\sigma /(2\vartheta))^2 \left( m_\sigma/\MP\right)^2,
\eeq
where $\kappa \equiv B^2/4\pi \ll 1$.  The integration range $\tilde{a}$ in \reflef{mxelm_15}) is thus re-expressed by using \reflef{mxelm_19a});
\beq\label{exeq_2a}
\tilde{a}=\eta a \sim \kappa\left( m_\sigma/(2\vartheta) \right)^2
\left( m_\sigma/\MP\right)^\gamma.
\eeq
We first consider $\tilde{a}$ only due to the uncertainty in
the incident angle $\delta\vartheta$ for a fixed $\omega^2$;
\beq
\tilde{a}=\omega^2 -\omega_r^2 \sim
-\frac{\partial\omega_r^2}{\partial\vartheta } \delta\vartheta
=\half\left(\frac{m_\sigma}{\vartheta} \right)^2
\frac{\delta\vartheta}{\vartheta}.
\label{exeq_2b}
\eeq
Combining this with \reflef{exeq_2a}) we obtain
\beqa
\label{exeq_3}
\delta\vartheta/\vartheta \sim (\kappa/2)
 \left( m_{\sigma}/\MP \right)^{\gamma}.
\eeqa
We emphasize that the resonance condition \reflef{exeq_1}) defines
not a point but a hyperbolic band in the $\vartheta-\omega$ plane 
given a finite $\delta\vartheta$. 
This implies that a deviation $\delta\omega$ from the nominal $\omega$ can
satisfy the same resonance condition with a different $\vartheta$ 
within $\pm \delta\vartheta$. 
As far as $\delta\omega/\omega \ll
\delta\vartheta/\vartheta$ is satisfied in a setup,
we may ignore the effect of $\delta\omega$.

%
%
We introduce an experimental resolution $\epsilon$ defined by 
$\epsilon \equiv |\delta\vartheta/\vartheta|$, giving
$\gamma \sim \ln(2\epsilon/\kappa)/\ln(m_{\sigma} /\MP)$, and hence
\beq
\eta 
\sim (M_P/m_{\sigma})^2 (2\epsilon/\kappa),
\label{exeq_4}
\eeq
due to \reflef{mxelm_19a}).  
Substituting this into \reflef{mxelm_19}), we derive
%
%
\beqa\label{exeq_5}
\overline{\frac{d\sigma}{d\Omega_3}}
\sim \frac{\pi}{8\omega^2}\vartheta^{-4}
\left( \frac{m_{\sigma}}{M_P} \right)^2
\left( \frac{\kappa}{2\epsilon} \right),
\eeqa
in the extremely forward direction. 
It is remarkable that the small $\vartheta \sim 10^{-9}$ produces 
a huge factor which nearly compensates
$(m_{\sigma}/M_P) \sim 10^{-36}$, leaving us with another $10^{-36}$
which can be taken care of by a sufficiently strong laser beam. 
In addition, thanks to the narrow forward peak, 
measuring $\omega_3 \sim 2\omega$ frees us
from measuring the angle $\theta_3$ directly to the demanding resolution
$\sim 10^{-9}$.

%
%
In principle we can explore the entire mass range $m_{\sigma}<\pi\omega$
by using two crossing beams with small incident angles.
Then we can directly measure the resonance curve in (\ref{exeq_1})
by scanning both $\vartheta$ and $\omega$ to observe the
resonance nature explicitly.  For the smaller mass scale such as
$m_{\sigma} \lsim 10^{-9}$~eV, however, we must take the finite
beam size due to the diffraction limit into account for controlling the small
incident angle with realizable optical devices and a distance on the ground.

\begin{figure}
\includegraphics[width=7.0cm]{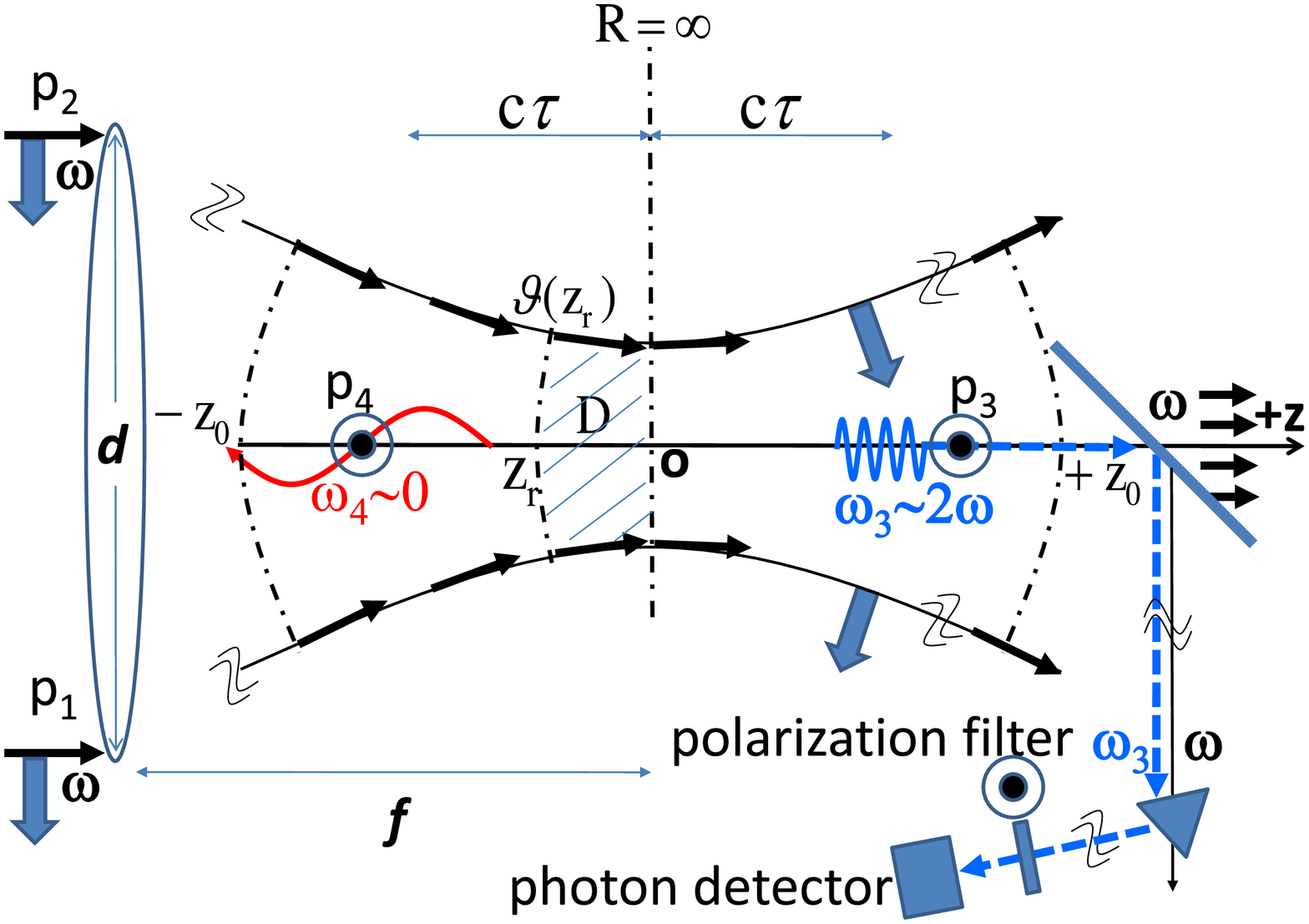}
\caption{
A conceptual experimental setup.
}
\label{Fig2}
\end{figure}

For this purpose, we now propose a conceptual experimental setup with
one-beam focusing  as illustrated in Fig.\ref{Fig2}. 
Incident photons from a Gaussian laser pulse linearly polarized 
to the state 11 are focused by the conceptual thin lens component 
into the diffraction limit with a reasonable focal 
length to satisfy the resonance condition.
The quasi-parallel incident photons interact with each other
around the focal point, from which photons 3 and 4 are emitted 
nearly in the opposite direction along the $z$ axis 
with $\omega_3\sim 2\omega$ and $\omega_4 \sim 0$.
The mirror with a dichroic nature is almost transparent to the 
non-interacting photons, while $\omega_3$ is reflected 
to the prism (equivalent to a group of dichroic mirrors)
which selects $\omega_3$ among residual $\omega$ 
and sends it to the photon detector placed off the $z$ axis. 
This process is assisted by the polarization filter 
selecting the rotated state 22. 

%
%

%
%
The electric field component in the Gaussian beam 
as a function of spatial coordinate $(x,y,z)$ is well-known \cite{Yariv},
\beqa\label{exeq_6}
E \!\propto\! \frac{w_0}{w(z)}\exp \hspace{-.2em}
\left(\hspace{-.2em}
-i\!\left[
\rule[-.1em]{0em}{1.0em}kz\!+\!\frac{kr^2}{2R(z)}-\phi(z)\right]
-\hspace{-.2em} \frac{r^2}{w(z)^2}
\right)\!,
\eeqa
where $k=2\pi/\lambda$ with wavelength $\lambda$,
$r=\sqrt{x^2+y^2}$, $w_0$ being the minimum waist, while other definitions are  
 curvature $R(z) = z (1+{z_0}^2/z^2)$, phase $\phi(z) = \tan^{-1}
 (1+z/z_0)$ and the waist $w(z) $ as a function of $z$;
\beqa\label{exeq_7}
{w(z)}^2 = {w_0}^2 ( 1+z^2/{z_0}^2),
\eeqa
with a parameter $z_0 \equiv \pi{w_0}^2/\lambda$.
%
%
We introduce the  $F$-number with the focal length $f$ and
the laser beam diameter $d$ defined by $F = 2f/(\pi d)$.
Then the beam waist at $z=0$ is given by $w_0 =
(d/2)(f/z_0)/\sqrt{1+(f/z_0)^2} \sim F \lambda$ for $f \ll z_0$ which is
the case we are interested in and we focus on the diffraction limit in
$|z| < z_0$ in what follows. 

According to \reflef{exeq_6}), curvature $R=\infty$ is
exactly satisfied at $z=0$ with $\vartheta = 0$.
We may thus expect the resonance condition (\ref{exeq_1})
to be met  automatically somewhere in the region $-z_0 < z < 0$.
The upper limit of the resonance angle $\vartheta(z_r)$ on the same
wavefront (equi-phase surface) can be estimated from \reflef{exeq_7})
 for $|z_r| \ll z_0$;
\beqa\label{exeq_8}
\vartheta(z_r) \equiv \frac{w(z_r)-w_0}{z_r}
\sim \frac{w_0 z_r}{2{z_0}^2} 
= \frac{1}{2\pi^2 F^3} \frac{z_r}{\lambda}.
\eeqa

%
%
At any $z_r$, the incident angle between any combinations of two
photons from the same wavefront varies between
$0< \vartheta < \vartheta(z_r)$. The mean $\vartheta$ of the two photons
chosen randomly from the above range originating from a spherical
wavefront is $\bar{\vartheta}\equiv (1/3)\vartheta(z_r)$.  It then follows that the experimental resolution cannot be $\epsilon \le 2$ for any $z_r$.
For $z_r \le z < 0$, called the domain $D$, on the other hand, the condition
$0<\vartheta(z)<\vartheta(z_r)$ is satisfied.  We then find that 
$\epsilon \sim {\cal O}(1)$ can be assigned for the domain $D$ as the
upper limit on the incident angular resolution.

Let us estimate effective luminosity in $D$.  Suppose a Gaussian
laser pulse with the duration time $\tau$ satisfying $c\tau \le z_0$
with the light velocity $c$ enters $D$ from the left side with the
average number of photons $\bar{N}$. The effective number of photons in $D$
which allows the use of the cross sections in \reflef{exeq_5})
with $\epsilon \sim {\cal O}(1)$ is defined by $\bar{N}_D =
|z_{r}|/(c\tau) \bar{N}$.  The effective luminosity per transverse area inside
the laser pulse in $D$ is expressed as
\beqa\label{eq_19}
{\cal L} = b\: C(\bar{N}_D,2)/\left(\pi w^2(z_r)\right) \sim
\left( \pi F\omega \bar{\vartheta}/c^2\tau \right)\bar{N}^2,
\eeqa
where $C(\bar{N}_D, 2) \approx \bar{N}_D^2/2$, while $b$ is for how many
domains with 
fixed $\epsilon$ are contained in the incident pulse with 
the total length $c\tau$, hence $b=c\tau/|z_r|$ with $z_r$
defined by $\vartheta(z_r)$ in \reflef{exeq_8}), with further approximation 
$w(z_r)\sim w_0$.

%
%
Multiplying \reflef{exeq_5}) times $c^2$ by \reflef{eq_19}), 
we obtain the differential yield \cite{footnote4},
\beqa\label{eq_20}
\frac{d{\cal Y}}{d\Omega_3}
= \frac{\pi^2 }{16} \bar{\vartheta}^{-3} (F/\omega \tau)
(m_{\sigma}/M_P)^2 (\kappa/\epsilon) \bar{N}^2,
\eeqa
per pulse rather than per unit time, since \reflef{eq_19}) 
includes the effect from the entire pulse. 
We then define $\bar{N}_1$ by 
\beqa\label{eq_21}
\bar{N}_1 = 
 \frac{4}{\pi}
\sqrt{\frac{ \epsilon \tau \omega \bar{\vartheta}^3}{\kappa F}}
(M_P/m_{\sigma}),
\eeqa
for $d{\cal Y}/d\Omega_3 = {\cal O}(1)$, or a single photon per pulse focusing.

%
%
Although this proposal applies both to CW and pulsed laser systems 
by optimizing $F/\omega\tau$ and $\bar{N}$, we here estimate a rate for a short laser pulse system based on \reflef{eq_21}) with
$\epsilon ={\cal O}(1)$ and the physical parameters: $\bar{\vartheta}
\sim 10^{-9},\: \omega \sim 1$~eV with $\kappa \sim 10^{-4}$ and 
$M_P \sim 10^{27}$~eV.  
For $F \sim 10^2$ and $c\tau \sim 1 ({\rm eV})^{-1}$ 
with $\tau \sim 4$~fs, we find $\bar{N}_1 = 10^{23} \sim 10$~kJ 
per pulse focusing \cite{footnote4}.
Since the conceptual lens component must have a reasonable aperture size 
to keep the incident power density below the damage threshold,
the dichroic mirror is assumed to be located 
at the symmetric position from the focal point at shortest.
The solid angle is then estimated to be $d\Omega_3 \sim F^{-2} \sim  10^{-4}$.
For a 10~Hz repetition rate 
the signal rate is $10^{-3}$~Hz assuming the perfect detection efficiency
for $\omega_3$ after the mirror. 

%
%
A major instrumental background for the doubled
frequency appears to come from the second harmonic generation~(SHG)
due to gas-solid interfaces with the centrosymmetry broken maximally.
Even from the maximal estimate $\sim10^{13}$W/cm${}^2$ for a typical damage
threshold, we find a negligible amount of
$10^{-10}$ SHG photons from a 1m${}^2$ aperture
size with a 10~fs irradiation, if the optical components are housed 
in a vacuum containing $10^{10}$~atoms/cm${}^3$ ($\sim 10^{-5}$~Pa)~\cite{SHG}.

%
%
As a dominant physical background we expect the lowest-order QED
photon-photon scattering, with the forward cross section,  $\sim
(\alpha^2/m_e^4)^2 \omega^6 \vartheta^4$ \cite{XsecQED}. 
This  turns out to be smaller than  \reflef{exeq_5}) by 50 orders of
magnitude  
for the above parameter values, indicating  the QED contribution 
to be totally negligible.
The resonance effects due to a pseudoscalar-field exchange from 
axion-like particles can also be suppressed if the initial 
photon polarizations are all in parallel as in our conceptual design. 

%
%
In view of these estimates of the suppressed backgrounds, instrumental and
physical, our proposal is expected to be a basis for realizing 
actual experiments by respecting the novel ideas in overwhelming 
the weak gravitational coupling by such non-gravitational effects 
like the small incident angle and the high laser intensity.

%
%
\begin{acknowledgments}
K. Homma thanks D. Habs, R. H\"{o}rline, S. Karsch, T. Tajima,
S. Tokita, L. Veisz and M. Zepf for valuable discussions.
This work was supported by the Grant-in-Aid for Scientific
Research no.21654035 from MEXT of Japan in part and the DFG Cluster of
Excellence MAP (Munich-Center for Advanced Photonics).
\end{acknowledgments}

%
%


\begin{references}
\bibitem{footnote2}We have $B=(1/3)(\alpha/\pi){\cal Z}\zeta$,
which is $2/3$ times as large as the coefficient in (6.194)
in \cite{cup}, because of slightly  different processes and
using half-spin fermions instead of spinless fields
in a toy model in the loop.
\bibitem{footnote3}The four digits in the subscript are for
$\beta$ arranged from left to right according to the photon labels, 1-4.
\bibitem{footnote4}We point out the potential importance of
the possible coherent amplification of the scattering amplitude over the entire
laser volume which is covered by the long force-range of the scalar field,
with likely a huge effect of multiplication by another factor $\bar{N}$
on the cross section \cite{SMITH_COHERENT}.
\end{references}
\end{document}